\documentclass[10pt,a4paper,twoside]{article}
\usepackage[english]{babel}
\usepackage[latin1]{inputenc}
\usepackage{amssymb}
\usepackage{mathrsfs}
\usepackage{amsmath}
\usepackage[dvips]{graphicx}
\usepackage{epsfig,bm}
\usepackage{graphicx}
\usepackage{rotate}
\usepackage{vmargin}

\usepackage{amsfonts}
\usepackage{graphics}
\usepackage{theorem}

\usepackage{color}
\usepackage{setspace}
\usepackage{amscd}
\usepackage{pstricks}
\usepackage{lscape}
\usepackage{url}
\usepackage{setspace}
\usepackage{harvard}

\title{Dynamical Complexity in the C.elegans Neural Network}
\author{Chris G. Antonopoulos$^{1}$, Athanasios S. Fokas$^{2}$ and Tassos C. Bountis$^{3}$}

\date{}

\begin{document}

\maketitle

\begin{center}
$^{1*}$Department of Mathematical Sciences, University of Essex, Wivenhoe Park, CO4 3SQ, United Kingdom
\\
$^{2}$Department of Applied Mathematics and Theoretical Physics, University of Cambridge, Cambridge, CB3 0WA, United Kingdom
\\
$^{3}$Center for Research and Applications of Nonlinear Systems, Department of Mathematics, University of Patras, Patras, GR-26500, Greece
\\
$^{*}$ Corresponding author email: canton@essex.ac.uk
\par\end{center}

\noindent\textbf{Summary}\\
We model the neuronal circuit of the \textit{C.elegans} soil worm in terms of Hindmarsh-Rose systems of ordinary differential equations, dividing its circuit into six communities pointed out by the walktrap and Louvain methods. Using the numerical solution of these equations, we analyze important measures of dynamical complexity, namely synchronicity, the largest Lyapunov exponent, and the $\Phi_{\mbox{AR}}$ auto-regressive integrated information theory measure, which has been suggested to reflect different levels of consciousness. We show that $\Phi_{\mbox{AR}}$ provides a useful measure of the information contained in the \textit{C.elegans} brain dynamic network. Our analysis reveals that the \textit{C.elegans} brain dynamic network generates \textit{more information than the sum of its constituent parts}, and that attains higher levels of integrated information for couplings for which either all its communities are highly synchronized, or there is a mixed state of highly synchronized and desynchronized communities. Both situations are characterized by relatively low chaotic behavior.



\section*{Introduction}\label{sec_intro}

Single-cell organisms manage to survive without possessing neurons. For example, bacteria in a Petri-dish respond to a drop of a toxic substance by clamping together. Presumably, neurons appeared in evolution when multicellular organisms were sufficiently complicated that it became ``useful'' to have a designated system of communication. Organisms with even very simple nervous systems exhibit more complex behaviors than organisms without neurons. For example, the \textit{C.elegans} soil worms, which have 302 neurons, feed alone if food is available and if the environment is quiet; however, if food is scarce and if they detect a threat (such as an unpleasant odour), they feed in groups. Presumably, this behavior is unconscious.

The identification of objective criteria for distinguishing conscious from unconscious processes still remains an important open problem. In the particular case of the human brain, leading neuroscientists in the area of consciousness, such as S. Dehaene and colleagues \cite{Dehaene2011200} have emphasized the following ``signatures of consciousness'': (i) The amplification of neuronal activity leading  to a sudden activation of parietal and prefrontal neuronal circuits, (ii) the appearance of a late slow wave in electroencephalographic (EEG) recordings (the so-called P3 wave), (iii) a late and sudden burst of high-frequency oscillations in EEG recordings, and (iv) the synchronization of activity in remote areas of the brain.

In recent years, an attempt has been made to ``mathematise'' consciousness \cite{Tononietal2003,Tononietal2004,Tononietal2005,Balduzzietal2008,Tononietal2008,Balduzzietal2009,Oizumietal2014}. Indeed, in his beautiful book ``Phi, a Voyage from the Brain to the Soul'' \cite{Tononi2012}, the well-known neuroscientist and psychiatrist G. Tononi claims that the only way to understand consciousness is to express it in terms of mathematical equations. Furthermore, Tononi claims that ``consciousness is integrated information theory'', and the latter takes indeed the form of a concrete mathematical expression. Another leading neuroscientist, C. Koch, in his recent book ``Consciousness'' \cite{Koch2012} goes even further: He claims that Tononi's notion of integrated information theory leads to ``a consciousness-meter that can assess the extent of awareness in animals, babies, sleepers, patients, and others who cannot talk about their experience''.

In what follows, we focus on the brain dynamic network (BDN) of the \textit{C.elegans} soil worm \cite{cmtkdataset} whose connectome is almost completely mapped \cite{Varshneyetal2011}. Our work primarily focuses on several quantities that describe the dynamical complexity of this brain network, and after computing these quantities, we make comparisons and draw a number of useful conclusions with respect to chaotic behavior, neural synchronization and a measure that quantifies the amount of integrated information generated by this network. In more detail, taking into consideration the importance of chaos \cite{Ott2002} in the general theory of complexity \cite{Bar-yam2003,Nicolisetal2007,Fuchs2013} and its connection to consciousness \cite{Tononietal1998}, as well as the role of synchronization \cite{Belykh_2005,Gardnesetal2010,Baptistaetal2010,Arenasetal2008,Mathias2011} and integrated information theory as signatures of consciousness \cite{Tononietal2004,Tononietal2008,Oizumietal2014}, we compute and compare measures of chaos \cite{Benettin1980}, synchronization \cite{Gardnesetal2010}, and integrated information theory \cite{Barrettetal2011} based on the numerical solution of a BDN modeling the \textit{C.elegans} brain.

It is well-known that a defining feature of all neural circuits (including the primitive radiata) is their connectivity. Obviously, the larger the number of neurons and the higher their connectivity, the richer the behavior of the associated neuronal circuit. Neuronal network modeling provides a rigorous mathematical way of quantifying this behavior. Indeed, building on the seminal work of Hodgkin and Huxley \cite{Hodgkinetal1952} there now exist several different systems of ordinary differential equations (ODEs) which can be used to model{ a given neural circuit. The numerical solution of these ODEs exhibits typical features of the behavior of a neuronal circuit, including chaotic behavior characterized by spiking and bursting.

\section*{Results}\label{sec_results}

Our analysis is based on the numerical solution of the ODEs \eqref{HR_model_Nneurons}. Using the walktrap method, we have divided the \textit{C.elegans} BDN into 6 communities (the Louvain method also produced 6 communities), and have computed the synchronization measure $\rho$ (see Eq. \ref{z_t} below), as well as the synchronization measures $\{\rho_{c_i}\}_1^6$ for each of the 6 communities using both methods. These parameters are plotted in panel (a) (for $\rho$) and panels (c)-(h) (for $\rho_{c_1}$-$\rho_{c_6}$) of Fig. \ref{fig_results_walktrap} as functions of the nonlinear coupling strength $g_n$ and the linear coupling strength $g_l$. The former characterizes the strength of the links between the different communities, whereas the latter the strength of the links within each community. These graphs show that for low nonlinear couplings and moderate to higher linear couplings, all communities, as well as the full network become highly synchronized with $\rho>0.9$ (depicted by the yellow and orange area in the parameter space). This corresponds to the case where the internal synapses within each community are stronger with respect to the synapses that link the different communities. However, this synchronization pattern starts to change as the nonlinear coupling increases to higher values. Interestingly, synchronization patterns start to emerge after $g_n>0.15$ and for high enough linear couplings: In this regime, the third and sixth communities become synchronized, whereas the other communities remain in a highly desynchronized state. This situation is reminiscent of the so-called ``chimera states'' that have been recently observed also in simple network models of coupled Hindmarsh-Rose (HR) oscillators, where synchronized and asynchronous populations are found to coexist \cite{HKB2014}.

The largest Lyapunov exponent $\lambda_1$ is depicted in panel b) of Fig. \ref{fig_results_walktrap}. This graph shows that $\lambda_1$ is rather high (red region) in the region of the parameter space where the synchronization levels remain quite low. Since higher values of $\lambda_1 (>0)$ are associated with a higher degree of chaos, this implies an ``inverse'' relation between synchronization levels and the level of dynamical instability (i.e. chaos) of neural activity. Of course, depending on the coupling strengths, there are also regions where both quantities are low.

We have computed the auto-regressive $\Phi_{\mbox{AR}}$ for both the membrane potential $p(t)$ and fast ion current $q(t)$. Indeed, for each pair of values in the plane of nonlinear and linear coupling parameters, the numerical simulation produces a recorded time-series $X^p(t)=\{p_1(t),\ldots,p_{277}(t)\}$ and $X^q(t)=\{q_1(t),\ldots,q_{277}(t)\}$ of the neural activity of all 277 neurons (for an explanation why we used 277 out of 302 neurons see Subsec. \ref{subsubsection_C.elegans_data} C.elegans Data). The \textit{C.elegans} brain network was divided into 6 unequally distributed communities. Thus, it is more convenient for the estimation of $\Phi_{\mbox{AR}}^p$ and $\Phi_{\mbox{AR}}^q$, to compute for each community $c_i,\;i=1,\ldots,6$, the following averaged time-series
\begin{equation}
X^p_{c_i}(t)=\frac{1}{|c_i|}\sum_{j\in c_i}X^p_j(t),
\end{equation}
and
\begin{equation}
X^q_{c_i}(t)=\frac{1}{|c_i|}\sum_{j\in c_i}X^q_j(t),
\end{equation}
where $|c_i|$ is the number of neurons in community $i$.

The next step is to prepare the community-averaged time-series in such a way as to be able to calculate $\Phi_{\mbox{AR}}^p$ and $\Phi_{\mbox{AR}}^q$, based on the averaged $X^p_{c_i}$ and $X^q_{c_i}$ versions of the data. In this respect, we transform $X^p_{c_i}$ and $X^q_{c_i}$ into stationary time-series by performing initially a detrending procedure (subtracting from them their mean), adjusting any difference in time of their variance by computing the logarithm with base 10 of $X^p_{c_i}$ and $X^q_{c_i}$, and removing the parts of the trajectories that correspond to quiescent periods (i.e. absence of spiking activity). In this way, we obtain the stationary versions $\bar{X}_{c_i}^p$ and $\bar{X}_{c_i}^q$, which can be seen as multivariate (six variates for each $\bar{X}_{c_i}$) analogues of stochastic-like random stationary processes. These quantities can now be used for the estimation of $\Phi_{\mbox{AR}}^p$ and $\Phi_{\mbox{AR}}^q$. 

For the core estimation of these two quantities we use the Matlab code ``ARphidata.m'' for stationary data provided in the Supporting Information of Ref. \cite{Barrettetal2011}. We thus present the results of these computations for $\tau=1$ (associated with the walktrap community detection method) in panels i), j) of Fig. \ref{fig_results_walktrap}. Panel i) is the parameter space for the quantity $\Phi_{\mbox{AR}}^p$ and panel j) for the quantity $\Phi_{\mbox{AR}}^q$. In the context of the integrated information theory of consciousness \cite{Tononietal2003,Tononietal2004,Tononietal2005,Tononietal2008}, $\Phi_{\mbox{AR}}\ge0$, being zero when a system generates the same amount of information with the sum of its parts as it transitions between states. If one is to attribute a physical meaning to $\Phi_{\mbox{AR}}^p$ and $\Phi_{\mbox{AR}}^q$, one would say that the higher their values, the higher the level of consciousness. In this sense, we cannot claim that a certain value of $\Phi$ is small, unless it is compared with other $\Phi$ values in the same figure. For this reason, we do not normalize $\Phi_{\mbox{AR}}^p$ and $\Phi_{\mbox{AR}}^q$. We observe that $\Phi_{\mbox{AR}}^p$ reaches high values in the range of low nonlinear and relatively moderate to high linear couplings; these are precisely the values where the synchronization levels of all communities (panels c) to h)) and the whole BDN (panel a)) of the \textit{C.elegans} are also quite high. On the other hand, $\Phi_{\mbox{AR}}^q$ attains high values in the range of high nonlinear and relatively low to high linear couplings; this is the region of parameter space where the synchronization level of the sixth (see panel h)) and possibly the third community (see panel e)) are very high, despite the asynchronous behavior of the rest of the communities.

A similar study for $\tau=1$ presented in Fig. \ref{fig_results_Louvain} for the six communities detected by the Louvain method reveals analogous results with those for the walktrap method shown in Fig. \ref{fig_results_walktrap}. In these two figures, there are, however, some striking differences with regard to the global synchronization $\rho$ and integrated information theory measure $\Phi_{\mbox{AR}}^p$. Indeed, a comparison between Figs. \ref{fig_results_walktrap}a) and \ref{fig_results_Louvain}a) shows that the whole BDN of the \textit{C.elegans} becomes highly synchronized (Fig. \ref{fig_results_Louvain}a)) for $g_n>0.25$, which is in contrast to what is depicted in Fig. \ref{fig_results_walktrap}a) for the same range of nonlinear couplings. The reason is that the number of nonlinear connections for the Louvain method is larger (742 undirected links) than those of the walktrap method (586 undirected links). Additionally, the integrated information theory measures $\Phi_{\mbox{AR}}^p$ in Figs. \ref{fig_results_walktrap}i) and \ref{fig_results_Louvain}i) attain their highest values (about 0.3 and 0.4 respectively) for different coupling ranges: Notably, $\Phi_{\mbox{AR}}^p$ is maximal in Fig. \ref{fig_results_Louvain}i) for $g_l$ values in $(1.6,2)$ and small nonlinear coupling $g_n$, whereas in Fig. \ref{fig_results_walktrap}i), it is maximal for pairs of couplings for which $g_l>0.8$ and $g_n$ quite small. On the other hand, when the linear coupling is moderate and the nonlinear coupling large enough (i.e. $g_n\approx0.28$), $\Phi_{\mbox{AR}}^q$ in Figs. \ref{fig_results_walktrap}j) and \ref{fig_results_Louvain}j) attains its highest value (of about 0.09 in both cases). It is well-known that, in the context of the integrated information theory of consciousness, a good practise is to adopt the $\tau$ value that maximizes $\Phi_{\mbox{AR}}$. Here, in order to study the effect of different time delays in $\Phi_{\mbox{AR}}$, we have also computed similar parameter spaces for both integrated information theory measures and community detection methods for $\tau=2,3,4$ and 5. Interestingly, we found that the new parameter spaces about both $\Phi$ measures for $\tau=2,3,4$ and 5 are qualitatively similar to those in Figs. \ref{fig_results_walktrap} and \ref{fig_results_Louvain} for $\tau=1$, and thus there is no need for choosing those that maximize $\Phi_{\mbox{AR}}$. For comparison, we present the case for $\tau=5$ in the Supplemental Figure S1 online. In conclusion, both measures attain higher levels for couplings for which either all communities are strongly synchronized, or there is a mixed state of highly synchronized and desynchronized communities. Both cases are found to be characterized by low chaotic behavior, as depicted by the maximal Lyapunov exponent.

\begin{figure}[!ht]
\centering{
\includegraphics[width=\textwidth,height=18cm]{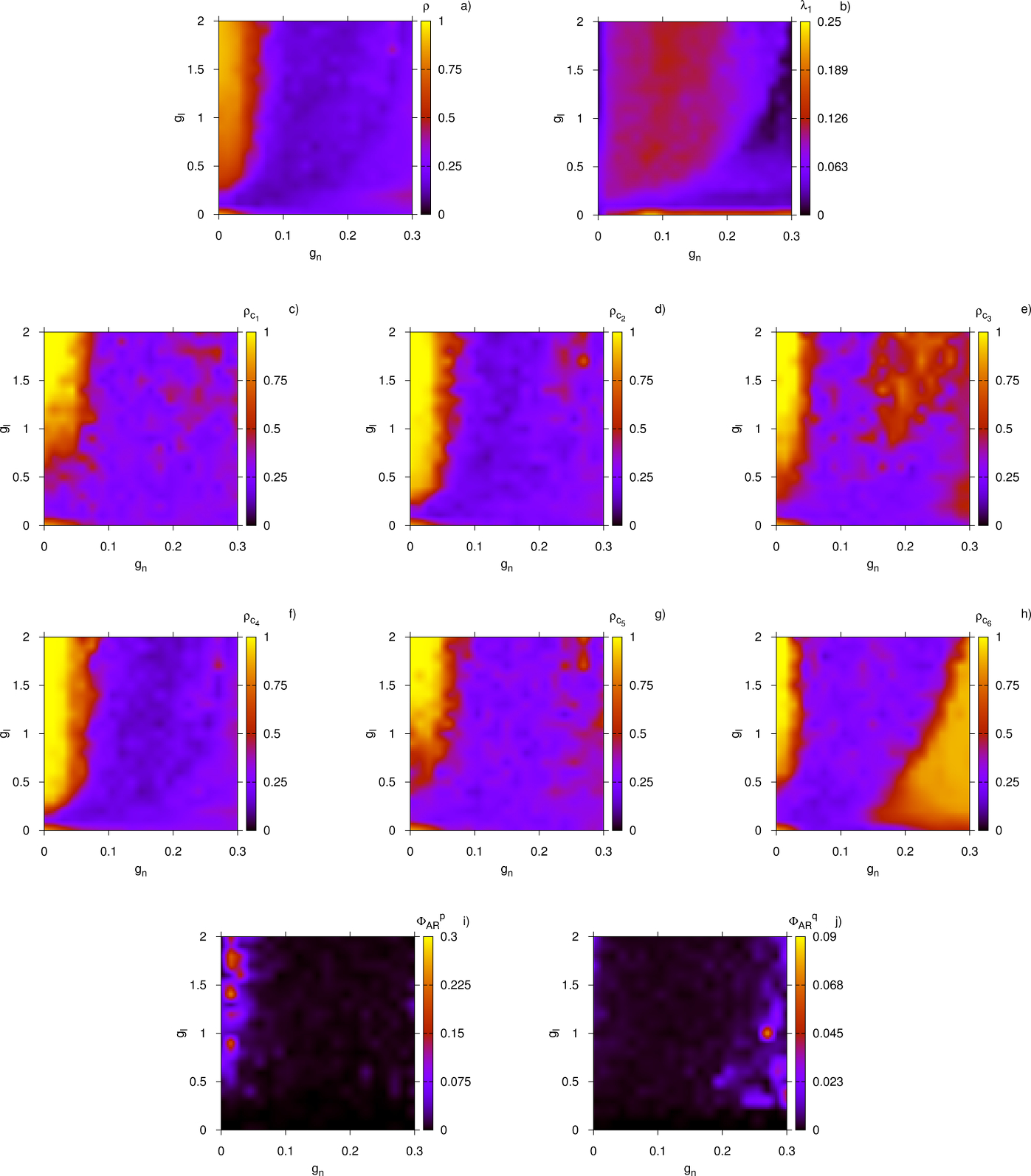}
}
\caption{High synchronization and low dynamical neural instability imply high integrated information theory measure of consciousness levels for the \textit{C.elegans} communities detected by the walktrap method. Panel a) is the parameter space of the synchronization measure $\rho$ for the whole BDN and panel b) is a similar parameter space for the largest Lyapunov exponent $\lambda_1$ of the neural dynamics. Panels c) to h) are similar to panel a) for the synchronization measures $\rho_{c_1}$-$\rho_{c_6}$ of the six communities, respectively. Panels i) and j) show the parameter spaces of the integrated information theory measures $\Phi_{\mbox{AR}}^p$ and $\Phi_{\mbox{AR}}^q$, respectively. In all panels, $g_n$ is the nonlinear coupling, $g_l$ the linear coupling, and $\tau=1$.}\label{fig_results_walktrap}
\end{figure}

\begin{figure}[!ht]
\centering{
\includegraphics[width=\textwidth,height=18cm]{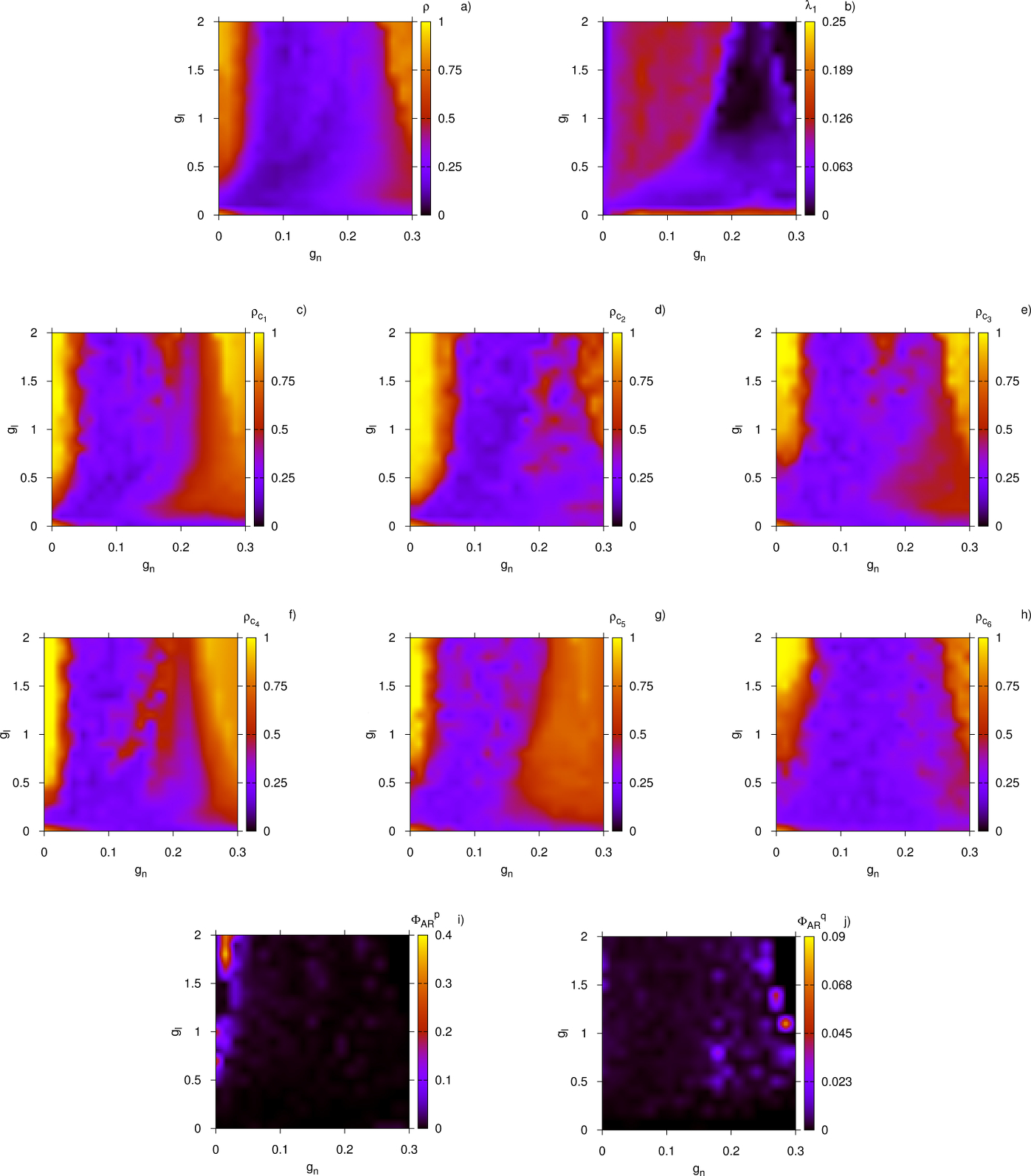}
}
\caption{High synchronization and low dynamical neural instability imply high integrated information theory measure of consciousness levels for the \textit{C.elegans} communities detected by the Louvain method. Panel a) is the parameter space of the synchronization measure $\rho$ for the whole BDN and panel b) is a similar parameter space for the largest Lyapunov exponent $\lambda_1$ of the neural dynamics. Panels c) to h) are similar to panel a) for the synchronization measures $\rho_{c_1}$-$\rho_{c_6}$ of the six communities, respectively. Panels i) and j) show the parameter spaces of the integrated information theory measures $\Phi_{\mbox{AR}}^p$ and $\Phi_{\mbox{AR}}^q$, respectively. In all panels, $g_n$ is the nonlinear coupling, $g_l$ the linear coupling, and $\tau=1$.}\label{fig_results_Louvain}
\end{figure}

\section*{Discussion}

The human brain has about 86 billion neurons \cite{CNE:CNE21974} whereas \textit{C.elegans} has only 302. However, it is well-known that evolution uses similar approaches for the solution of similar problems, and thus features of the neuronal dynamics of \textit{C.elegans} may have more general validity. In this respect, it should be noted that both the human and the \textit{C.elegans} nervous systems consist of neurons that communicate either chemically via special ``synapses'', or linearly via ``gap junctions''. Chemical communication is achieved via the so-called neurotransmitters. Many of these neurotransmitters are common in humans and \textit{C.elegans}, including glutamate, GABA, acetylcholine and dopamine. The genome of the \textit{C.elegans} is almost 30 times smaller than that of humans, but still encodes almost 22000 proteins; moreover, it is about 35\% similar to that of humans \cite{Blumenthal1996,mcdgicelegans}. Electrical communication is bidirectional and of local character, i.e. it exists between neurons whose cells are close. They are believed to contribute to the regulation of synchronization in the brain network.

Each conscious state consists of a myriad of different parts and is distinct from any other experience. At the same time, it is experienced as a coherent whole that is absolutely impossible to be separated into parts. This union of ``differentiation'' and ``integration'' is a fundamental property of consciousness. Perhaps the dynamical complexity of neural systems is directly related to the above fundamental property.

In this work we have attempted to quantify and compare certain important measures of dynamical complexity. In particular: (i) A measure of synchronicity, which is merely suggestive of integration between sub-domains as the system exhibits coherent behavior as a whole (see Ref. \cite{Srinivasanetal1999} in which the authors show that when a simple flickering stimulus is consciously perceived there is a marked increase in long-range coherence at the stimulus frequency), (ii) the largest Lyapunov exponent, which provides a measure of chaos (or dynamical instability) and, (iii) $\Phi_{\mbox{AR}}^p$ and $\Phi_{\mbox{AR}}^q$, which are the auto-regressive information measures associated with the membrane potential $p$ and fast ion current $q$ respectively, and can be understood as measures of the extent to which the present global state of the system reflects the past global state, compared with predictions based on the most informative decomposition of the system into its components. 

It appears that the notion of $\Phi_{\mbox{AR}}$ provides a useful tool for quantifying the integrated information contained in a given neural system. In the context of the integrated information theory of consciousness, if one is to attribute a physical meaning to $\Phi_{\mbox{AR}}$, one would say that the higher its value, the higher the level of consciousness.

Combining the results of Figs. \ref{fig_results_walktrap} and \ref{fig_results_Louvain}, and based on the above interpretation, our analysis suggests that, for particular coupling strengths, the \textit{C.elegans} BDN is able to generate more information than the sum of its constituent parts. Specifically, we found that the \textit{C.elegans} BDN attains higher levels of integrated information for couplings for which either all its communities are highly synchronized, or there is a mixed state of highly synchronized and desynchronized communities, a situation that corresponds to low chaotic neural behavior. We found that in the case of the \textit{C.elegans} brain network there exist substantial differences between the behaviors of the $\Phi_{\mbox{AR}}^p$ and $\Phi_{\mbox{AR}}^q$ measures.

A complementary approach has been given in Ref. \cite{Antonopoulosetal2015} where various statistical quantities associated with the \textit{C.elegans} brain network, such as the global clustering coefficient, the average of local clustering coefficients, the mean shortest path, the degree probability distribution function of the network and the small-worldness measure have been computed. Even though small-worldness captures important aspects of complex networks at the local and global scale, it does not provide information about the intermediate scale. This can be better described by the modularity, or community structure of the network.

Finally, it is important to note that in Ref. \cite{Varshneyetal2011} the authors have gathered and combined material from many different sources and studies, and reported on the whole set of self-consistent gap junction and chemical synapses of the \textit{C.elegans} brain. They identified neurons that may play a central role in information processing, and network motifs that could serve as functional modules of the brain network. This is in the same spirit with the notion of communities used in the present study, but involves a more complicated distribution of chemical and electrical synapses than we have assumed here. In a future publication, we plan to extend our analysis to investigate dynamical complexity in more ``realistic'' neural networks, such as those reported in Ref. \cite{Varshneyetal2011}.

\section*{Experimental Procedures}\label{online_methods}

\subsection*{\textit{C.elegans} Data}\label{subsubsection_C.elegans_data}

\textit{C.elegans} is a 1mm long soil worm with a simple nervous system that consists of 302 neurons and about 7000 synapses \cite{Gallyetal2003}. Its nervous system is divided into two distinct and independent components: A large somatic nervous system and a small pharyngeal nervous system. In our study we use the connectome of the large somatic nervous system found in Ref. \cite{cmtkdataset}, which consists of 277 neurons. In the present work we are not interested in and do not investigate the important questions of the directionality of the information flow, thus we use the undirected version of the relevant adjacency matrix. We couple the different neurons via ODEs using the corresponding adjacency matrix obtained from the brain connectivity of the \textit{C.elegans}.

\subsection*{The Hindmarsh-Rose Neural Model}\label{subsection_Hindmarsh-Rose_Model_for_Brain_Dynamics}

We model the dynamics of each ``neuron'' by a single HR neuron system. Namely, following Refs. \cite{Baptistaetal2010,Antonopoulosetal2015}, we endow the nodes (i.e. neurons) of the \textit{C.elegans} BDN with the dynamics characterized by the following system of ODEs \cite{Hindmarshetal1984}
\begin{eqnarray}\label{HR_model_1neuron}
 \dot{p}=q-ap^3+bp^2-n+I_{\mbox{ext}}\nonumber,\\
 \dot{q}=c-dp^2-q\nonumber,\\
 \dot{n}=r[s(p-p_0)-n],
 \end{eqnarray}
where $p$ is the membrane potential, $q$ characterizes the fast ion current (i.e. $Na^{+}$ or $K^{+}$), and $n$ the slow ion (adaptation) current, for example $Ca^{2+}$. In this neuron system, $p$, $q$ and $n$ are expressed in dimensionless units \cite{Hindmarshetal1984}. The parameters $a$, $b$, $c$, $d$, which model the function of the fast ion channels, and $s$, $p_0$ are given by $a=1$, $b=3$, $c=1$, $d=5$, $s=4$ and $p_0=-8/5$ (see Ref. \cite{Hindmarshetal1984}) and are the same for all neurons. Parameter $r$, which modulates the slow ion channels of the system, is set to $0.005$ for all neurons, and the parameter $I_{\mbox{ext}}$, which is the current that enters the neuron, is set to 3.25 for all neurons. For the above values, each neuron can exhibit chaotic behavior and the solution for $p(t)$ exhibits typical multi-scale chaos characterized by spiking and bursting activity which is consistent with the membrane potential observed in experiments on single neurons \textit{in vitro} \cite{Hindmarshetal1984}. Thus, chaos not only allows the simulated BDNs to reproduce behaviors empirically observed in experiments with single neurons, but also provides an interesting mechanism for the network to process information generated via an external stimulus \cite{Baptistaetal2008A}.

We couple the HR system to create an undirected BDN of $N_n$ neurons connected simultaneously by linear diffusive coupling and nonlinear coupling synapses \cite{Antonopoulosetal2015}
\begin{eqnarray}\label{HR_model_Nneurons}
 \dot{p}_i=q_i-a p_i^3+bp_i^2-n_i+I_{\mbox{ext}}-g_n(p_i-V_{\mbox{syn}})\sum_{j=1}^{N_n}\mathbf{B}_{ij}S(p_j)-g_l\sum_{j=1}^{N_n}\mathbf{G}_{ij}H(p_j)\nonumber,\\
 \dot{q}_i=c-dp_i^2-q_i\nonumber,\\
 \dot{n}_i=r[s(p_i-p_0)-n_i],\nonumber\\
 \dot{\phi}_i=\frac{\dot{q}_i p_i-\dot{p}_i q_i}{p_i^2+q_i^2},\;i=1,\ldots,N_n,
\end{eqnarray}
where $\dot{\phi}_i$ is the instantaneous angular frequency of the $i$-th neuron, with $\phi_i$ being the phase defined by the fast variables $(p_i,q_i)$ of the $i$-th neuron \cite{Pereiraetal2007}. The functions $H$ and $S$ are chosen as $H(p_i)=p_i$ and $S(p_j)=[1+e^{-\lambda(p_j-\theta_{\mbox{syn}}}]^{-1}$ \cite{Baptistaetal2010}. $S$ is a sigmoidal function that acts as a continuous mechanism for the activation and deactivation of the chemical synapses and, also allows for analytical calculations of the synchronous modes and synchronization manifolds of the coupled system of Eqs. \eqref{HR_model_Nneurons} \cite{Baptistaetal2010}. The remaining parameters are chosen as follows: For the parameters $\theta_{\mbox{syn}}$, $\lambda$, and $V_{\mbox{syn}}$, we set $\theta_{\mbox{syn}}=-0.25$, $\lambda=10$, and $V_{\mbox{syn}}=2$ is chosen so as to yield an excitatory BDN. 

The parameters $g_n$ and $g_l$, which are varied in the parameter spaces of all figures of the paper, denote the strength associated with the nonlinear excitatory and linear diffusive coupling between the corresponding synapses. Regarding the physical meaning of these parameters, we note that the authors in Ref. \cite{Varshneyetal2011} have suggested that there exist chemical and self-consistent gap junction synapse networks in the \textit{C.elegans} brain. In our model, we have used a simplified version, in which neurons in different communities are coupled chemically (through nonlinear excitatory connections) whereas those in the same community are coupled electrically (through linear diffusive connections). It would be interesting to extend our study to more complex brain networks such as those described in Ref. \cite{Varshneyetal2011}.

For the chosen parameters, we have taken $|p_i|<2$, and $(p_i-V_{\mbox{syn}})$ always negative for excitatory networks. If two neurons are connected via an excitatory synapse, then if the presynaptic neuron spikes, it induces the postsynaptic neuron to spike. We adopt only excitatory nonlinear synapses in our analysis. $\mathbf{G}_{ij}$ accounts for the way neurons are linearly (diffusively) coupled and is represented by a Laplacian matrix
\begin{equation}
\mathbf{G}_{ij}=\mathbf{K}_{ij}-\mathbf{A}_{ij},
\end{equation}
where $\mathbf{A}$ is the binary adjacency matrix of the linear connections and $\mathbf{K}$ is the degree identity matrix based on $\mathbf{A}$; thus $\sum_{j=1}^{N_n}\mathbf{G}_{ij}=0$. By binary we mean that if there is a connection between two neurons then the entry of the matrix is 1, otherwise it is 0. $\mathbf{B}_{ij}$ is a binary adjacency matrix and describes how the neurons are nonlinearly connected and therefore its diagonal elements are equal to 0, thus $\sum_{j=1}^{N_n}\mathbf{B}_{ij}=k_i$, where $k_i$ is the degree of the $i$-th neuron, i.e. $k_i$ represents the number of nonlinear links that neuron $i$ receives from all other $j$ neurons in the network. A positive off-diagonal value of both matrices in row $i$ and column $j$ means that neuron $i$ perturbs neuron $j$ with an intensity given by $g_l\mathbf{G}_{ij}$ (linear diffusive coupling) or by $g_n\mathbf{B}_{ij}$ (nonlinear excitatory coupling). Therefore, the binary adjacency matrix $\mathbf{C}$ of the complex networks considered in this work is given by
\begin{equation}
\mathbf{C}=\mathbf{A}+\mathbf{B}.
\end{equation}
We use as initial conditions for each neuron $i$ the following: $p_i=-1.30784489+\eta^r_i$, $q_i=-7.32183132+\eta^r_i$, $n_i=3.35299859+\eta^r_i$ and $\phi_i=0$, where $\eta^r_i$ is a uniformly distributed random number in $[0,0.5]$ for all $i=1,\ldots,N_n$ (see Ref. \cite{Antonopoulosetal2015} for details). With these initial conditions the trajectory converges very quickly to the attractor of the dynamics and thus, there is less need to consider longer transients. 

\subsection*{Numerical Simulations Details}\label{subsection_numerical_simulation_details}

We have integrated numerically Eqs. \eqref{HR_model_Nneurons} using the Euler integration method (first order) with time step $\delta t=0.01$. We have decided to employ a first order scheme in order to reduce the numerical complexity and CPU time of the required simulations to feasible levels. A preliminary comparison of the trajectories computed for the same parameters (i.e. $\delta t$, initial conditions, etc.) via integration methods of order 2, 3 and 4 produced similar results.

We have calculated the largest Lyapunov exponent $\lambda_1$ using the well-known method of Ref. \cite{Benettin1980}. The numerical integration of the HR system of Eqs. \eqref{HR_model_Nneurons} was performed for a total time of $t_f=5000$ units and the computation of the various quantities needed in our analysis, such as the largest Lyapunov exponent $\lambda_1$, were computed after a transient time $t_t=300$ in order to make sure that orbits have converged to an attractor of the dynamics.

\subsection*{Synchronization Measures in BDNs}\label{subsection_a_measure_of_synchronization_in_brain_networks}

It is known that burst synchronization of neural systems can be strongly influenced by many factors, including coupling strengths and types \cite{Belykh_2005}, noise \cite{Buric_2007}, and the existence of clusters in neural networks \cite{Viana_2014}. Here, we use the order parameter $\rho$ to account for the synchronization level of the neural activity of the \textit{C.elegans} BDN and its communities \cite{Gardnesetal2010}. This notion, which originates from the theory of measures of dynamical coherence of a population of $N_n$ oscillators of the Kuramoto type \cite{Kuramotoetal_2002}, can be computed via the expression \cite{Gardnesetal2010}
\begin{equation}\label{z_t}
 z(t)=\rho(t)e^{\mathrm{i}\Psi(t)}=\sum_{j=1}^{N_n}e^{\mathrm{i} \phi_j(t)},
\end{equation}
where $N_n$ denotes the number of neurons of the BDN and $\phi_j(t)$ is the phase variable of the $j$-th neuron of the HR system given by the fourth equation in \eqref{HR_model_Nneurons}. The modulus $\rho(t)$ of the complex number $z(t)$, which takes values in $[0,1]$, measures the phase coherence of the population of the $N_n$ neurons, and $\Psi(t)$ measures the average phase of the population of oscillators. Actually, we average $\rho(t)$ over time to obtain the order parameter $\rho\equiv\langle\rho(t)\rangle_t$, which determines the tendency of $\rho$ in time. The value $\rho=1$ implies complete synchronization of the oscillators, whereas $\rho=0$ means complete desynchronization.

We use Eq. \eqref{z_t}, adapted accordingly, for the computation of the synchronization level of the \textit{C.elegans} BDN, and of its communities (for a discussion of communities see Analysis of Networks and Communities subsection). In particular, $N_n$ is the number of neurons of the BDN and $j$ runs through all $N_n=277$ neurons of that network, whereas in the case of communities, $N_n$ represents the number of neurons of the particular community and $j$ refers to the particular neurons which are members of this community.

\subsection*{Analysis of Networks and Communities}\label{subsection_analysis_of_communities}

\subsubsection*{\textit{C.elegans} Brain Network}

We identified the communities of the \textit{C.elegans} brain network using two different approaches: The walktrap method \cite{Ponsetal2005} which employs the igraph software using six steps \cite{Antonopoulosetal2015}, and the Louvain method \cite{Blondeletal2008} (with resolution 1) which employs the NetworkX software \cite{networkx}.

The walktrap algorithm detects communities using a series of short random walks based on the idea that vertices encountered on any given random walk are more likely to lie within a community. The algorithm initially treats each node as its own community, and then merges them into larger communities, followed by still larger ones, and so on. Essentially, given a graph, this algorithm tries to find densely connected subgraphs (i.e. communities) via random walks. The idea is that short random walks tend to stay in the same community. Using the above procedure we have been able to identify 6 communities in the \textit{C.elegans} brain network.

The Louvain algorithm involves two phases: In the first phase, by optimizing modularity locally, it looks for ``small'' communities, and in the second phase the algorithm aggregates nodes of the same community and builds a new network whose nodes are the communities. These steps are repeated iteratively until a maximum of modularity is achieved. By focusing on ad-hoc networks with known community structure, it has been shown that the Louvain method is very accurate. Moreover, due to its hierarchical form which is reminiscent of renormalization strategies, this method allows one to look for communities at different resolutions. The output therefore yields several partitions: The partition found after the first step typically consists of many communities of small sizes; at subsequent steps, larger and larger communities are found due to the aggregation mechanism, and this process naturally leads to a hierarchical decomposition of the network. This algorithm is obviously an approximate method and nothing ensures that the global maximum of modularity is attained, but several tests have confirmed that the Louvain algorithm is quite accurate and often provides a decomposition into communities with a modularity close to the optimal. The Louvain algorithm outperforms other methods in terms of computation time, and this allows one to analyze networks of unprecedented size \cite{Blondeletal2008}.

Following this procedure we were able to identify again 6 communities in the \textit{C.elegans} brain network. However, the communities identified by the walktrap and Louvain methods are not identical, neither in size nor in their members.

\subsection*{Integrated Information Theory Measures}\label{subsection_integrated_information_measures}

In Ref. \cite{Barrettetal2011} the authors present some practical methods for measuring integrated information \cite{Tononietal2004} from time-series data. Based on recently introduced measures of {\it integrated information} (see for example Refs. \cite{Tononietal2003,Balduzzietal2008,Oizumietal2014}), they analyze quantities that measure the extent to which a system generates more information than the sum of its constituent parts as it transitions between different states. These measures possibly reflect levels of consciousness generated by neural systems. The authors in Ref. \cite{Barrettetal2011} propose two new such measures, $\Phi_{\mbox{E}}$ (empirical) and $\Phi_{\mbox{AR}}$ (auto-regressive), that overcome limitations faced by older versions of analogous quantities, and can be computed using time-series data derived from measurements of realistic or model systems. Thus, these measures offer promising approaches for revealing relations between integrated information, consciousness, and other neurocognitive quantities in real and model systems.

The auto-regressive $\Phi$ ($\Phi_{\mbox{AR}}$) is well-suited for cases where the time-series is non-Gaussian distributed but nevertheless stationary and stochastic. By construction, when applied to Gaussian-distributed, stationary data, it is equivalent to the well-known empirical version of $\Phi$ for integrated information, $\Phi_{\mbox{E}}$. However, these measures already differ when applied to non-Gaussian, stationary data. Indeed, $\Phi_{\mbox{AR}}$ provides a useful measure of integrated information based on relations between conditional entropy, partial covariance and linear regression prediction error.

Next, we briefly describe the derivation of $\Phi_{\mbox{AR}}$ following Ref. \cite{Barrettetal2011}: We start from two multivariate random variables $X=(X^1,\ldots,X^n)^T$ and $Y=(Y^1,\ldots,Y^m)^T$, where $^T$ denotes the transpose of a matrix or a vector. Their linear regression is then given by
\begin{equation}\label{linear_regression}
X=a+A\cdot Y+E,
\end{equation}
where $A$ is the regression matrix, $a$ is a vector of constants, and $E$ the prediction error. $E$ is a random vector uncorrelated with $Y$. Given that the distributions of $X$ and $Y$ are defined by
\begin{eqnarray}
A&=&\sum(X,Y)\sum(Y)^{-1},\\
a&=&\bar{x}-A\cdot \bar{y},
\end{eqnarray}
it follows that the linear regression is unique. In this framework, $(Y)^{-1}$ denotes the inverse of the covariance matrix of $Y$, $\sum(X)$ denotes the $n\times n$ matrix of covariances, $\sum(X,Y)$ the $n\times m$ matrix of cross-covariances, and $\bar{x}$, $\bar{y}$ are the means of the random variables $X$ and $Y$ respectively. $E$ has zero mean and its covariance is the partial covariance of $X$ given $Y$,
\begin{equation}\label{sum_E}
\sum(E)=\sum(X|Y)=\sum(X)-\sum(X,Y)\sum(Y)^{-1}\sum(X,Y)^T.
\end{equation}
Provided that $\sum(Y)$ is an invertible covariance matrix, Eq. \eqref{sum_E} holds for any random variables $X$ and $Y$, whether they are Gaussian or not.

If it happens that $X$ and $Y$ are Gaussian distributed random multivariate variables, then the conditional entropy of $X$ given that $Y=y,\;y\in\mathbb{R}^m$ satisfies the equation
\begin{equation}
H(X|Y=y)=\frac{1}{2}\log\Bigl(\det\Bigl(\sum(E)\Bigr)\Bigr)+\frac{1}{2}n\log(2\pi e),\;\forall y\in\mathbb{R}^m,
\end{equation}
where $\det()$ denotes the determinant of the matrix. This is a relation between the conditional entropy and linear regression prediction error valid for Gaussian systems. Under these assumptions, the effective information $\phi$ can be written as
\begin{equation}\label{phi_definition}
\phi\Bigl[X;\tau,\{M^1,M^2\}\Bigr]=\frac{1}{2}\log\biggl(\frac{\det(\sum(X))}{\det(\sum(E^X))}\biggr)-\frac{1}{2}\sum_{k=1}^2\log\biggl(\frac{\det(\sum(M^k))}{\det(\sum(E^{M^k}))}\biggr),
\end{equation}
where $E^{M^k},\;k=1,2$, and $E^X$ are the prediction errors in the linear regressions
\begin{eqnarray}
M_{t-\tau}^{k}&=&A^{M^k}\cdot M_t^k+E_t^{M^k},\\
X_{t-\tau}&=&A^X\cdot X_t+E_t^X.
\end{eqnarray}
Here, the notation $X_{t-\tau}$ denotes the $\tau$ steps past state (i.e. time lag) from the current state $X_t$. 

If the system under consideration is not Gaussian distributed, then Eq. \eqref{phi_definition} does not hold. However, the right hand side of Eq. \eqref{phi_definition} is a quantity that is well defined and can be measured empirically. This quantity is actually the basis of the alternative measure $\Phi_{\mbox{AR}}$, i.e. the auto-regressive measure $\Phi$ for integrated information proposed in Ref. \cite{Barrettetal2011}.

Summarizing, we assume that $X$ is a stationary, not necessarily Gaussian, multivariate random variable, and let $\phi_{\mbox{AR}}\Bigl[X;\tau,\{M^1,M^2\}\Bigr]$ represent the right hand side of Eq. \eqref{phi_definition}. Then, $\Phi_{\mbox{AR}}$ is simply $\phi_{\mbox{AR}}$ for the bipartition $\mathcal{B}=\{M^1,M^2\}$ of $X$ that minimizes $\phi_{\mbox{AR}}$ divided by the normalization factor
\begin{equation}
L(\mathcal{B})=\frac{1}{2}\log\Biggl(\min_k\Bigl\{(2\pi e)^{|M^k|}\det\Bigl(\sum\bigl(M^k\bigr)\Bigr)\Bigr\}\Biggr).
\end{equation}
Under these considerations, $\Phi_{\mbox{AR}}\bigl[X;\tau\bigr]$ is defined by
\begin{equation}\label{phiARcapital}
\Phi_{\mbox{AR}}\bigl[X;\tau\bigr]=\phi_{\mbox{AR}}\bigl[X;\tau,\mathcal{B}^{\mbox{min}}(\tau)\bigr],
\end{equation}
where
\begin{equation}
\mathcal{B}^{\mbox{min}}(\tau)=\arg_{\mathcal{B}}\min\Biggr\{\frac{\phi_{\mbox{AR}}[X;\tau,\mathcal{B}]}{L(\mathcal{B})}\Biggr\}.
\end{equation}

The function $\Phi_{\mbox{AR}}$, defined by Eq. \eqref{phiARcapital}, is formulated in terms of the linear regression prediction error, which essentially compares the whole system to the sum of its parts in terms of the log-ratio of the variance of the past state to the variance of the prediction errors of the linear regression of the past on the present. It can be understood as a measure of the extent to which the present global state of a system is predicted by the past global state, as compared to predictions based on the most informative decomposition of the system into its component parts. In other words, it is a measure that quantifies the extent to which a system generates more information than the sum of its constituent parts. Thus, as argued in Ref. \cite{Barrettetal2011}, $\Phi_{\mbox{AR}}$ possibly reflects levels of consciousness generated by neural systems.

It is important to note that in this work we have computed $\Phi$ using a \textit{macroscopic} partition of the associated network, as explained in the Results section. For the human brain it is an unproven hypothesis that macro-level $\Phi$ results correlate with micro-level $\Phi$ values. The interpretation of our results for the \textit{C.elegans} brain, therefore, with respect to integrated information, is based on a similar hypothesis.



\section*{Acknowledgements}
A. F. acknowledges support from EPSRC. We would like to thank Dr. N. Kouvaris for many fruitful discussions and for providing us the communities and adjacency matrices based on the Louvain community detection method. C. G. A., A. F. and T. B. acknowledge that this research has been co-financed by the European Union (European Social Fund - ESF) and Greek national funds through the Operational Program ``Education and Lifelong Learning'' of the National Strategic Reference Framework (NSRF) - Research Funding Program: THALES - Investing in knowledge society through the European Social Fund.


\end{document}